\documentclass[10pt,twocolumn,letterpaper]{article}
\usepackage[pagenumbers]{cvpr}

\definecolor{cvprblue}{rgb}{0.21,0.49,0.74}
\usepackage[breaklinks,colorlinks,allcolors=cvprblue]{hyperref}
\usepackage{pifont}
\usepackage{subcaption}
\usepackage{multirow}
\usepackage{float}
\newcommand{\Name}{BMO\xspace}

\title{Mind Your Boundaries: Boundary-Aware Model Obfuscation against Cross-node Attack}
\author{Yuyang Zhang\textsuperscript{*}\quad
Yihui Jin\textsuperscript{*}\quad
Kangjie Chen\quad
Leyang Wu\quad Run Wang\\[0.5em]
{\normalsize \textsuperscript{*}Equal contribution.}
}
\hypersetup{
  pdftitle={Mind Your Boundaries: Boundary-Aware Model Obfuscation against Cross-node Attack},
  pdfauthor={Yuyang Zhang, Yihui Jin, Kangjie Chen, Leyang Wu, Run Wang},
  pdfsubject={Preprint}
}

\begin{document}
\maketitle
\begin{abstract}

With the advancement of powerful AI chips, on-device machine learning is gaining traction for its low latency and privacy protection. 
However, when edge devices are compromised, attackers can obtain white-box access to the deployed models by reversing the apps directly to obtain the knowledge of peer devices, which is known as \textit{Cross-Node Attacks}, like performing efficient adversarial attacks. 
Most existing defenses target known vulnerabilities, yet white-box access still enables attackers to uncover more unknown weaknesses.
This creates a critical risk for edge–device deployments, where one compromised device can endanger all models distributed from the same source. 
Motivated by findings that the model robustness is primarily influenced by the curvature of the decision boundary, we investigate how model parameters shape its curvature.
In this paper, we propose a novel method, \Name, that obfuscates model parameters to diversify the decision boundary further resisting cross-node attack while preserving its benign accuracy. 
Specifically, we explore decomposing model parameters into boundary-position and boundary-curvature components based on null-space constraints, then models with differentiated decision boundaries are instantiated by injecting noise into the boundary-curvature component.
Across all four major types of adversarial attacks with four widely-deployed on-device models, \Name achieves an average success rate of 68.68\% in defense, far exceeding the most competitive baseline, which attains only 22.32\% in defending against cross-node attacks. 
Moreover, \Name achieves comparable performance in benign accuracy compared with baselines.

\end{abstract}
\section{Introduction}

In recent decades, deep learning (DL) models have been extensively employed across various cutting-edge fields (\eg, image classification, speech recognition and natural language processing), achieving remarkable and transformative progress~\cite{ding2022roadmap}.
More recently, powered by the highly efficient AI chips and model compression technologies, on-device DL models have been developed and widely assembled into edge devices, such as mobiles~\cite{singh2023edge}, to deliver low‐latency inference and preserve user privacy to avoid potential leakage~\cite{murshed2021machine}. Such edge‐based deployments underpin a growing trend of real‐time, privacy‐sensitive services for end users across various application scenarios.

\begin{figure}[t]
    \centering
    \includegraphics[width=\linewidth]{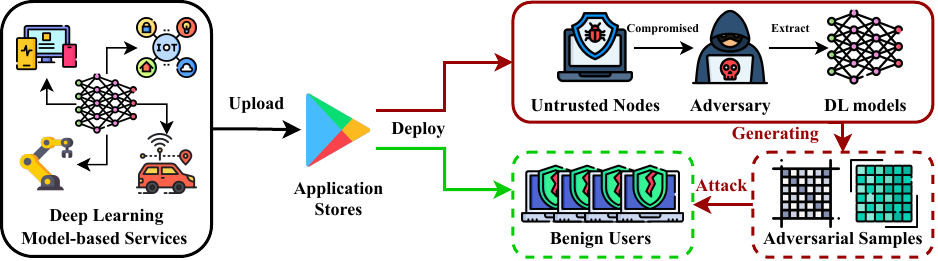}
    \caption{\textbf{An Illustration of \textit{Cross-node Attack}.} 
    End users install applications with integrated on-device models from app stores to enjoy the provided services. 
    Meanwhile, an adversary can exploit these applications by unpacking them to access the model in white-box settings. Then, the peer devices installed the same model are totally exposed to adversary leading a large-scale compromising.}
   
    \vspace{-5pt}
    
    \label{fig:scenario}
\end{figure}

Despite their benefits, recent studies have revealed that edge devices often operate in untrusted environments. In such settings, an adversary who compromises a single device can extract its installed on-device models~\cite{zhou2023nnsplitter}, gaining white-box access.  
Once obtained, the models distributed from the same source as the compromised model are totally exposed to the adversary, which enables efficient attacks, such as crafting effective adversarial samples targeting compromised models that show strong transferability among the peer devices~\cite{sun2025obfuscation}.
Therefore, in this work, we term this threat as the \textit{Cross-Node Attack}, exploiting full model internals under white-box conditions, unlike the well-known transfer-based adversarial attack refers to conducting efficient adversarial attacks under black-box scenarios.
As presented in Figure~\ref{fig:scenario}, such attacks typically involve two phases~\cite{zhou2024investigating}: \ding{182} extracting the on-device model from a compromised node, and \ding{183} conducting an efficient attack by leveraging the stolen parameters.

Existing studies show potential in defending the aforementioned cross-node attack, including \textit{model enhancement} and model \textit{asset protection}.
Model enhancement (\eg, adversarial training~\cite{wang2024revisiting}, input random~\cite{byun2022effectiveness}) aims to either improve its inherent capability by learning malicious samples or detecting samples passively with known anomalous patterns. Unfortunately, both of them are dedicated to known adversarial attacks~\cite{kumar2025fortifying}.
In contrast, model asset protection (\eg, model verification~\cite{zhou2024model}, obfuscation~\cite{zhou2023modelobfuscator}) investigates how to improve the costs of model extraction, further avoiding the white-box access to the victim models. 
However, recent studies reveal that such defenses offer no protection once the model is loaded into RAM on compromised nodes, where the parameters remain accessible and can be easily reversed or extracted through memory dumping or dynamic analysis~\cite{ijcai2022p94}.
Consequently, neither model enhancement nor asset protection truly suffices against cross-node attacks that obtain persistent white-box access and uncover unknown vulnerabilities, thereby rendering these approaches unreliable as practical defenses in this scenario~\cite{shen2024transferability}.

Motivated by the evidence that decision-boundary curvature is central to model robustness, we examine how model parameters shape this curvature~\cite{moosavi2019robustness,fawzi2016robustness,yang2020boundary}. 
Our preliminary experiment results show that existing defenses overlook the protection of boundary curvature, leaving models susceptible to cross-node attack. Once a curvature is exposed in one model, others with the same curvature can be compromised easily. 
To address this, we propose \textbf{B}oundary-aware \textbf{M}odel \textbf{O}bfuscation (\Name), which obfuscates decision-boundary curvature to address cross-node attacks while preserving model functionality well.
Specifically, inspired by \textit{Null-Space Constrained}~\cite{schwab2019deep}, \Name begins by decomposing the pre-trained weights into (i) \textit{magnitude components}, responsible for the prediction of the normal samples off the decision boundary, and (ii) \textit{direction components}, which shape the model's decision boundary. 
We then inject controllable noises \textit{solely} into the direction component to yield models whose decision-boundary curvature differs substantially, whereas the magnitude components ensure clean-input predictions remain intact.
This obfuscation requires only a single one-shot computation with negligible cost, unlike baselines that incur additional training or computational overhead.
Consequently, \Name substantially reduces the effectiveness of transferring a successful attack from the compromised node to other peer nodes, preventing the further propagation of malicious behavior within the network.

We evaluate \Name across all 4 major types of adversarial attacks and 4 widely deployed on-device models. 
Our experimental result indicates that \Name effectively resists attacks generated by diverse adversarial techniques, reaching an average 68.68\% success rate in cross-node defense while preserving 97.54\% benign accuracy  ($\Delta=-1.08\%$), substantially outperforming the best baseline, which attains only 22.32\%.
In addition, \Name remains robust under model compression (\eg, quantization), sustaining a 59.63\% cross-node defense success rate with 97.54\% benign accuracy ($\Delta=-1.08\%$). 
To summarize, our key contributions are as follows:
\begin{itemize}[noitemsep,topsep=0pt,leftmargin=*]

\item We first reveal on-device models are vulnerable to cross-node attack, further present the characterization, detailing attacker capabilities, from extraction to adversarial payload generation, and quantifying transferability risks.

\item We introduce a novel defense paradigm for edge systems, \Name, which is a newly designed weight obfuscation strategy to generate device-specific models with distinct decision boundaries, significantly reducing the possibility in compromising other peer devices.

\item We introduce a novel defense that decomposes model weights into magnitude (preserving clean accuracy) and direction (shaping decision boundaries) components, where controlled noise are injected into the direction only to effectively create device-specific boundary shifts.

\item We integrate \Name into both convolutional and transformer pipelines with minimal overhead, and across four strong attacks and four network architectures  demonstrate our method exceeding the strongest baseline by 44\% in effectively defending against cross-node attack, while maintaining less than a 1\% drop in benign accuracy.

\end{itemize}
\section{Related Work}\label{sec:relatedworks}

\noindent\textbf{Adversarial attack.}
Adversarial attacks exploit imperceptible perturbations to manipulate the target models' prediction~\cite{liang2022adversarial}. Even minor, human-invisible modifications can lead to severe consequences, especially in security-sensitive applications such as autonomous driving~\cite{song2018physical,lovisotto2021slap}.
Depending on the attack conditions, adversarial attacks can be grouped into four categories:
\ding{182} gradient-based attacks (\eg, FGSM, PGD), which directly utilize model gradients to craft perturbations;
\ding{183} optimization-based attacks (\eg, C\&W), which solve constrained optimization problems to produce high-confidence adversarial examples;
\ding{184} query-based attacks (\eg, MI-FGSM), which rely on repeated queries to the model to estimate gradient information; and
\ding{185} decision-boundary–based attacks (\eg, HopSkipJumpAttack~\cite{chen2020hopskipjumpattack}), which approximate and exploit the decision boundary in a black-box manner.
In general, the more detailed knowledge an attacker possesses about the target model  and its internal parameters, the more effective their adversarial attack becomes in practice~\cite{liang2022adversarial}.

However, in cross-node scenarios, adversarial attacks pose a more serious threat and are more difficult to prevent since the identical model is deployed across devices, compromising a single untrusted node grants adversaries comprehensive knowledge of the same model and its defenses on other nodes. Unlike black-box attacks, where adversaries only approximate the victim, the cross-node adversary can craft perturbations with complete access to the internal gradients of the targeted model, significantly increasing the overall attack potency in realistic  scenarios~\cite{li2025enhancing}.

\noindent\textbf{Model defense.}
Primarily, there are two popular categories of model defense mechanisms employed on edge devices, \ie, model asset protection and model enhancement.
The model asset protection~\cite{ryffel2019partially} includes model structure obfuscation, modifying network architectures~\cite{zhou2023modelobfuscator} or replacing them with semantically equivalent code~\cite{zhou2024model}. 
However, recent works show that once the model is loaded into RAM on a compromised node, its parameters remain exposed and can be easily extracted via memory dumping or dynamic analysis~\cite{ijcai2022p94}.
Consequently, these defenses fail to provide protection in cross-node settings.
The model enhancement provides  defense against adversarial attacks through adversarial training~\cite{goodfellow2014explaining}, randomized architectures~\cite{yang2023simple,he2019parametric}, and interaction-level perturbations~\cite{athalye2018obfuscated}.
Unfortunately, despite promising developments, they  give only a “false sense of security”, as Athalye~\etal~\cite{athalye2018obfuscated} demonstrate, adaptive adversaries can  invert or average out such defenses.

Instead of focusing on enhancing model robustness or raising the cost of model white-boxing, our approach addresses this critical gap by strategically and efficiently employing a novel weight obfuscation to generate highly distinct models with low overlap of decision boundaries.

\section{Threat Model}

We consider a realistic scenario (Figure~\ref{fig:scenario}) in which the model owner packages a single trained model into an application and distributes the same binary to all users. Such deployments typically rely only on authentication for access control, without any per-user parameter differences~\cite{zeng2024implementation}.
We adhere to established scenarios from previous studies on extraction techniques~\cite{zhou2023modelobfuscator,zhou2024model} and adversarial attacks~\cite{madry2018towards,chen2020hopskipjumpattack} against edge AI applications.
The application package file (\eg{}, Apk file for Android, and IPA for iOS) can be decompiled by the off-the-shelf reverse engineering tools (\eg{}, Apktool, and IDA Pro) to get the original DL model files, which provides the potential for the adversaries to extract the internal information of models via various model extracting techniques. 
For example, the model $\mathcal{M}$ could be extracted by calling the system APIs of the development format directly~\cite{zhou2024model}.
Then, the adversary $\mathcal{A}$ opens the white-box access to the target on-device models, and the mobile device is compromised further. 
Different from transfer-based adversarial attacks, the cross-node attacks occur in the white-box condition, where the attackers have full information about the target models, which makes defense more challenging.

\subsection{Attacker's Goal}
After obtaining full knowledge of the on-device model deployed on the compromised nodes, the adversary could launch large-scale and powerful adversarial attacks (\eg{}, evasion attack) targeting peer nodes that embed the same DL applications.
This is formulated as follows:
\begin{equation}\label{eq:adversarialgoal}
\begin{array}{c}
\setlength\abovedisplayskip{3pt}%
\setlength\belowdisplayskip{3pt}
\mathcal{M}^{\prime}\left(x^{\prime}\right) = y, \quad \text { while } \quad\left\|x^{\prime}-x\right\| \leq \epsilon, \\
x^{\prime} = x+\delta, \quad \text { where } \delta = \min \mathcal{L}\left(\mathcal{M}(x+\delta), y\right)
\end{array}
\end{equation}
Here, $\epsilon$ controls the perturbation budget and the $\delta$ is the perturbation. Since the adversary does not directly access $\mathcal{M}^{\prime}$ at inference time, they optimize $x'$ via the extracted model $\mathcal{M}$, relying on the transferability property. $\mathcal{L}$ is the loss function used to mislead the model, and $y$ is the target output. The adversary expects that the adversarial input $x'$ will also mislead $\mathcal{M}$ despite not directly optimizing against it.

\subsection{Attacker's Knowledge and Capability}
\vspace{-5pt}
In a more practical scenario, we assume a ``weak attacker'' that the adversaries can access the DL apps through compromised nodes but lack any information about the source DL model. 
Additionally, since the adversaries may infer the original parameters by collecting massive edge models to fit the decision boundaries of the original model, we assume that a ``strong attacker'' can gain full knowledge of the original model boundaries. 
Both of the attackers can independently generate adversarial inputs with white- and black-box attack methods as well.

\subsection{Defender's Capability and Goal}
\vspace{-5pt}
As the owners of the models, vendors obviously do not have the ability to retrain a model for every software download, but they can still modify the models' architecture and training data at training time to enhance their robustness or obfuscate the models by modifying their parameters before distributing them to individual edge devices. 

The defenders aim to ensure that adversarial samples generated on one node are incapable of compromising DL models on other nodes, while maintaining predictive performance for benign samples.
Therefore, for defense, the goals of the vendor are as follows:
\begin{equation}
\setlength\abovedisplayskip{3pt}%
\setlength\belowdisplayskip{3pt}
\begin{array}{c}
    \min\mathcal{L}\left(\mathcal{M}^{\prime}(x), \mathcal{M}(x)\right),\\
    \max _{\|\delta\| \leq \epsilon} \mathcal{L}\left(\mathcal{M}^{\prime}(x+\delta), \mathcal{M}(x+\delta)\right)
\end{array}
\end{equation}
\section{Methodology}

\subsection{Model Weight Obfuscation}

Our key insight in defending cross-node attack is inspired by the \textit{Null-Space Constrained}~\cite{schwab2019deep} and is grounded in a fine-grained analysis of model weights. Specifically, we decompose the weight matrix into two complementary components: 
\textit{magnitude components} that determine the global position of the decision boundary and govern the model’s predictions for normal samples lying off the boundary,
and 
\textit{direction components} that control the local curvature of the boundary and encode features responsible for adversarial vulnerability. 
This formulation allows us to simultaneously achieve two objectives: \ding{182} maintain benign accuracy before and after model perturbation, and \ding{183} maximize the geometric discrepancy of decision boundaries, thereby reducing the risk of cross-node attack.
We further validate this hypothesis in Section~\ref{sec:experiments} via extensive and carefully designed experiments on multiple architectures and attack algorithms.

\begin{figure*}[t]
    \centering
    \includegraphics[width=\linewidth]{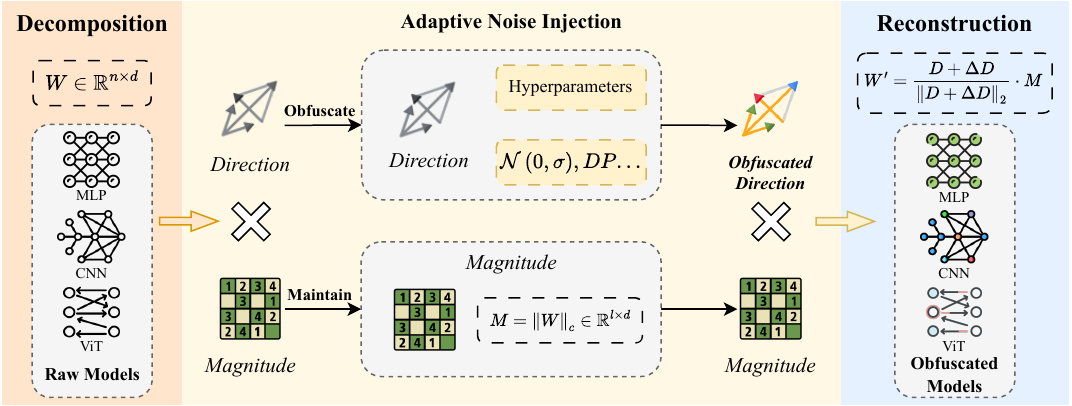}
    \caption{\textbf{An Overview of \Name.} \Name decomposes model weights into \textit{Magnitude} components related to the prediction of the normal sample, and \textit{Direction} components related to the shape of the decision boundary. It then obfuscates the model decision boundary by altering the direction components to reduce the effectiveness of cross-node adversarial samples, while maintaining the magnitude components to preserve predictive performance.}
    \label{fig:Methodology}
    
\end{figure*}

Building on this analysis, as shown in Figure~\ref{fig:Methodology}, we propose \textbf{B}oundary-aware \textbf{M}odel \textbf{O}bfuscation, \textbf{\Name}, which begins by splitting weights into a \textit{Magnitude} matrix as performance-aware components and a \textit{Direction} matrix as boundary-aware components.
Subsequently, we introduce controllable perturbations, such as Gaussian or differential privacy noise, to the direction component while maintaining the magnitude component constant. 
Notably, this obfuscation is performed only once at deployment and requires no retraining or additional procedures, thereby incurring negligible additional cost.

This approach injects controlled randomness into the subspace orthogonal to the model’s original capabilities, enabling the construction of diverse decision boundaries while preserving its core performance, in line with the null-space intuition.
The boundary diversity suggests that adversarial examples crafted on a compromised model are unlikely to be reproducible on other nodes, whose models exhibit different curvature features induced by independent direction perturbations.
Consequently, the attacker's knowledge about the compromised model cannot be reused on other nodes, effectively reducing the feasibility of cross-node attacks. This approach enhances the robustness of distributed models or applications, particularly in federated or decentralized settings, where weight-sharing and model replication are common.

\noindent\textbf{Orthogonal decomposition of weights.} 
Building upon these hypotheses, the foundation of our method lies in decomposing weights into direction and magnitude components, 
isolating perturbation-sensitive elements while retaining performance-aware parameters. Formally, for layer $\mathit{l}$ with weight $W^{(l)} \in \mathbb{R}^{n \times d}$, we calculate the Euclidean norm of each vector $W_{i, \cdot}^{(l)}$:
\begin{equation}\label{eq:decomposition}
\setlength\abovedisplayskip{3pt}%
\setlength\belowdisplayskip{3pt}
m_i = \left\| W_{i, \cdot}^{(l)} \right\|_2, \quad d_i = W_{i, \cdot}^{(l)}/(m_i + \epsilon) \quad (i = 1, \ldots, n)
\end{equation}
Collecting $m_i$ into a diagonal matrix $M^{(l)}$ and stacking the unit vectors $d_i$ into a matrix $D^{(l)}$:
\begin{equation}
\setlength\abovedisplayskip{3pt}
\setlength\belowdisplayskip{3pt}
\begin{aligned}
W^{(l)} &=  M^{(l)} \odot  D^{(l)} , \\
D^{(l)} &= \frac{W^{(l)}}{\|W^{(l)}\|_2 + \epsilon}, \\
M^{(l)} &= \mathrm{diag}(\|W^{(l)}_1\|_2, \|W^{(l)}_2\|_2, \ldots, \|W^{(l)}_m\|_2)
\end{aligned}
\end{equation}
where $\epsilon$ prevents division by zero. Similar to Dora~\cite{liu2024dora}, $M^{(l)}$ captures the original weight magnitudes and $D^{(l)}$ captures the directional structure, effectively separating magnitude from direction. We also demonstrated this hypothesis through the ablation studies in Figure~\ref{fig:Ablation}\subref{fig:WeightComponent}.

\noindent\textbf{Adaptive noise injection.}
After the decomposition of weights, we introduce noise into the disentangled weight components, generating multiple models with distinct decision boundaries, reducing decision boundary overlap across model instances. The goal of our weight obfuscation is to ensure functional consistency on benign inputs while inducing diversity among models, such that adversarial examples crafted on one model instance fail to transfer to others, effectively mitigating cross-node attacks. Concretely, we inject noise (\eg, Gaussian or Differential Privacy noise) into the direction component, which governs the shape of decision boundaries. Meanwhile, by preserving the magnitude matrix $M^{(l)}$,  we maintain the scale of activations, thereby retaining classification accuracy on benign samples.
In $D^{(l)}$, large value updates influence the decision boundary, so perturbing these directions significantly impacts its shape~\cite{liu2024dora}.
To minimize the effectiveness of adversarial attacks while minimizing performance loss, we inject noise into the top‑$p$ parameters of direction components. The selective randomization undermines an attacker’s ability to fit a static surrogate or compute reliable gradients, while maintaining nearly identical performance on benign data. 
For each directional matrix $D^{(l)}$, we identify critical elements using a sparsity-aware threshold where parameters exceeding this threshold are perturbed:
\begin{equation}
\setlength\abovedisplayskip{3pt}%
\setlength\belowdisplayskip{3pt}
    \text{Threshold}^{(l)}(p) = \text{Quantile}\left(|D^{(l)}|, 1-p\right),
\end{equation}
where $p \in [0,1]$ controls the proportion of perturbed elements(\eg, $p = 0.1$ perturbs the top 10\% of elements by absolute value). We then sample independent Gaussian noise scaled by $\alpha$:
\begin{equation*}
\setlength\abovedisplayskip{3pt}%
\setlength\belowdisplayskip{3pt}
    \Delta D_{i,j}^{(l)} = \alpha \cdot \mathcal{N}\left(0, \sigma^{(l)}\right) \cdot \mathbb{I}\left(\left|D_{i,j}^{(l)}\right| \geq \text{Threshold}^{(l)}(p)\right),
\end{equation*}
where $\sigma^{(l)}$ matches the standard deviation of $D^{(l)}$.

\noindent\textbf{Weight Reconstruction.}
We renormalize perturbed direction matrices and combine them with magnitude matrices to reconstruct the weights as Equation~(\ref{eq:reconstruction}). Strict normalization avoids the magnitude component $M^{(l)}$ being affected by noise. 
For each row, we perturb only the direction component and then renormalize it. Specifically, let
\begin{equation*}
\setlength\abovedisplayskip{3pt}%
\setlength\belowdisplayskip{3pt}
\tilde d_i = \frac{d_i + \Delta d_i}{\|d_i + \Delta d_i\|_2 + \epsilon},\quad i = 1,\ldots,n,
\end{equation*}
and denote by $\tilde D^{(l)}$ the matrix whose rows are $\tilde d_i$. The obfuscated weights are then reconstructed as
\begin{equation}\label{eq:reconstruction}
\setlength\abovedisplayskip{3pt}%
\setlength\belowdisplayskip{3pt}
W'^{(l)} = M^{(l)} \tilde D^{(l)}.
\end{equation}
This reparameterization magnitudes are preserved and only the normalized directions are perturbed.

\subsection{Gradient-based Decision Boundary Analysis}
\Name perturbs the direction components of pre-trained weights while preserving their magnitude components, introducing controlled randomness into the model’s decision boundary, which is a critical target in most adversarial attacks~\cite{he2018decision}. 
This process mimics finetuning on various independent datasets~\cite{hong2023publishing}, promoting boundary diversity across model instances, while preserving benign performance.
Prior work~\cite{fawzi2018empirical} highlights the role of decision boundaries in optimize the gradient descent directions, showing that the gradient at any point situated on decision boundaries is orthogonal to the tangent space. 
By altering the decision boundaries, \Name disrupts the optimization paths used in adversarial attacks~\cite{liu2024dora}, leading to divergence in adversarial perturbations across models. 
As a result, the direction of gradient descent used to optimize adversarial examples diverges across different model instances. Even when starting from the same input, it leads to significant differences in the generated adversarial examples, thereby reducing their transferability.
According to Equation~(\ref{eq:adversarialgoal}), adversaries typically craft perturbations $\delta$ via:
\begin{equation}
\setlength\abovedisplayskip{3pt}%
\setlength\belowdisplayskip{3pt}
    \delta^* = \arg\min_{\|\delta\| \leq \epsilon} \mathcal{L}(\mathcal{M}(x + \delta; \theta), y),
\end{equation}
where $\theta$ includes all $W^{(l)}$ across the network, $\mathcal{L}$ denotes the loss function (\eg, cross-entropy) and $y$ is the target label. Therefore adversarial attacks gain $\nabla_x \mathcal{L}$ and the derivative $\frac{\partial \mathcal{L}}{\partial W^{(l)}}$ following:
\begin{equation}
\setlength\abovedisplayskip{3pt}%
\setlength\belowdisplayskip{3pt}
    \nabla_x \mathcal{L} = \frac{\partial \mathcal{L}}{\partial W^{(l)}} \cdot \frac{\partial W^{(l)}}{\partial x} \xrightarrow{}   \frac{\partial \mathcal{L}}{\partial W^{(l)}} = \frac{\partial \mathcal{L}}{\partial D^{(l)}} \cdot \frac{\partial D^{(l)}}{\partial W^{(l)}}.
\end{equation}
and since $D^{(l)}$ is normalized per row, any noise $\Delta D^{(l)}$ perturbs the angular direction of gradients.

To further analyze the impact on gradient alignment, consider a small adversarial perturbation $\delta$ with sign aligned to $\nabla_x \mathcal{L}$. After applying \Name, the perturbed gradient becomes:
\begin{equation}
\setlength\abovedisplayskip{3pt}%
\setlength\belowdisplayskip{3pt}
    \widetilde{\nabla}_x \mathcal{L} = \nabla_x \mathcal{L} + \nabla_x \mathcal{L}_{\text{noise}}.
\end{equation}
The perturbation term $\nabla_x \mathcal{L}_{\text{noise}}$ introduces an angular deviation from the direction of the clean gradient. 
The perturbed gradient significantly misleads the adversary's optimization trajectory, thereby preventing the effective location of the decision boundary or even the gradient direction of the target model when using the extracted models.

Meanwhile, since the magnitude matrix $M^{(l)}$ remains unchanged, the inference capacity of the model during normal training are preserved. Therefore, our novel defense allows us to degrade gradient quality only for adversarial agents while maintaining the performance on benign data.

\begin{figure*}[t]
    \centering
    \begin{subfigure}{0.32\textwidth}
        \centering
        \includegraphics[height=5cm,width=\textwidth,keepaspectratio]{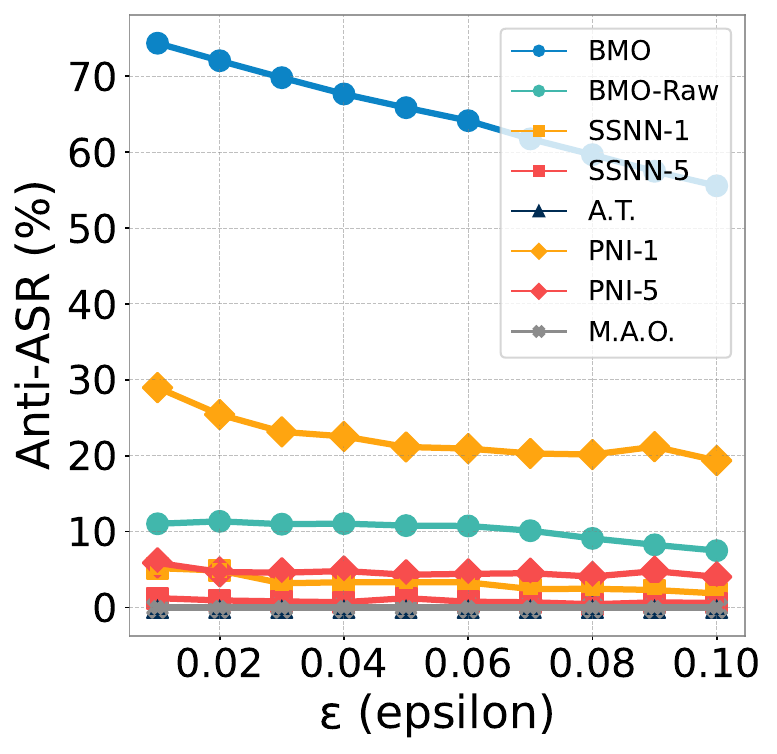}
        \caption{\textbf{Effectiveness with Attack Intensities.}}
        \label{fig:epsilon}
    \end{subfigure}
    \hfill
    \begin{subfigure}{0.32\textwidth}
        \centering
        \includegraphics[height=5cm,width=\textwidth,keepaspectratio]{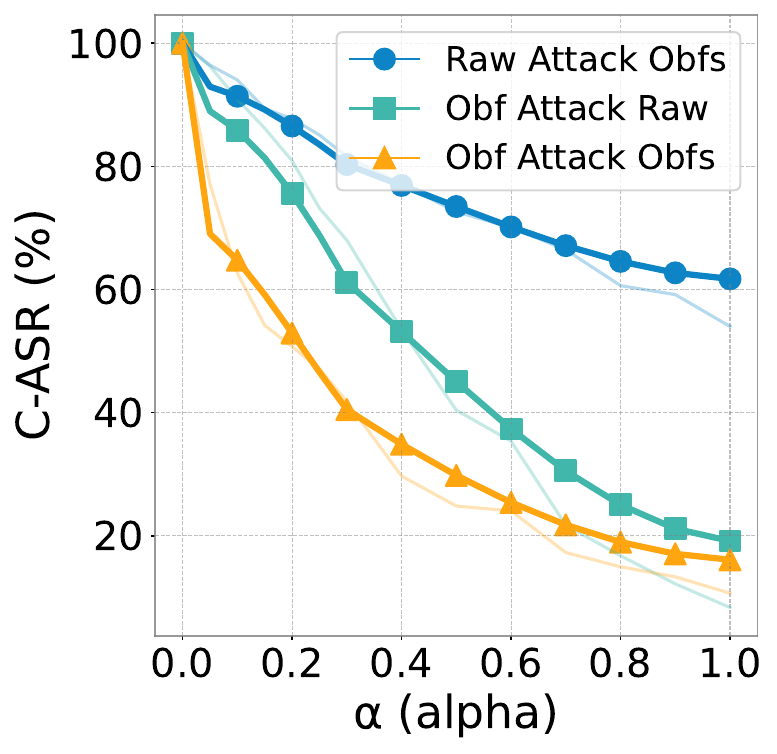}
        \caption{\textbf{Knowledge and Targets of Attackers.}}
        \label{fig:strict_knowledge}
    \end{subfigure}
    \hfill
    \begin{subfigure}{0.32\textwidth}
        \centering
        \includegraphics[height=5cm,width=\textwidth,keepaspectratio]{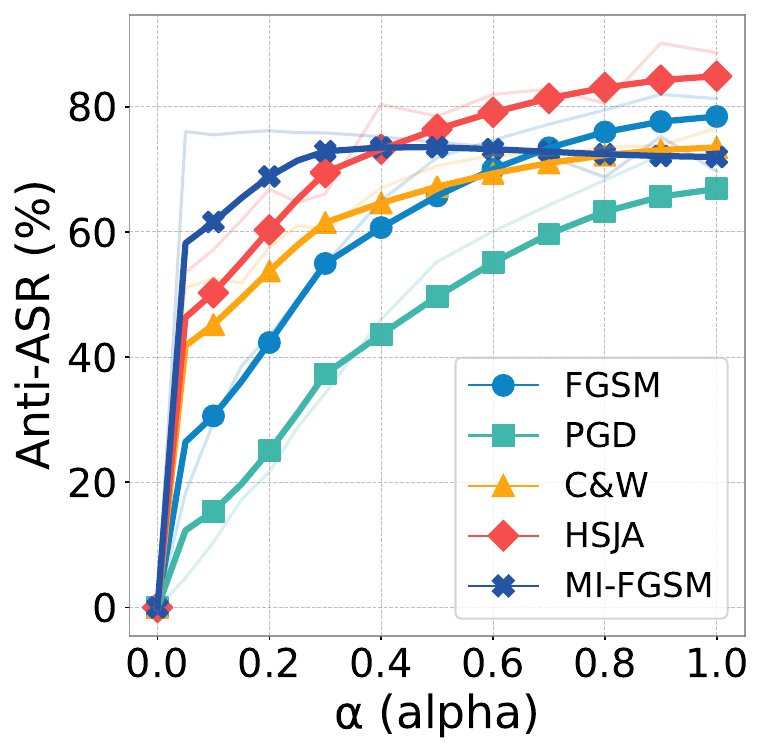}
        \caption{\textbf{Impact of Attack Methods.}}
        \label{fig:strict_attack_method}
    \end{subfigure}
    \caption{\textbf{Main Results of \Name.} We comprehensively evaluate the effectiveness of \Name under defense conditions by thoroughly moderating the three elements that determine the effectiveness of an attack, \ie, the intensities, algorithms, knowledge, and the attack targets.}
    \label{fig:main}
    \vspace{-10pt}
\end{figure*}

\section{Experiments}
\label{sec:experiments}

\subsection{Setups}

\noindent\textbf{Datasets and models.} 
To comprehensively assess the generalization of \Name, we evaluated it across three widely deployed model types on edge devices, \ie, fully connected neural networks~\cite{scabini2023structure}, convolutional neural networks~\cite{li2021survey}, and vision transformers~\cite{han2022survey}. Specifically, we employ the MLP, AlexNet, ResNet~\cite{he2016deep}, and ViT~\cite{dosovitskiy2020image} as typical models. Furthermore, experiments are conducted systematically and rigorously on three widely recognized datasets, \ie, MNIST~\cite{deng2012mnist}, CIFAR-10~\cite{krizhevsky2009learning}, and ImageNet~\cite{deng2009imagenet}.

\noindent\textbf{Adversarial attacks.} 
We conducted experiments on both white-box and black-box attack methods. For white-box attacks, we evaluated our method against the one-step FGSM attack~\cite{goodfellow2014explaining}, the multi-step PGD attack~\cite{madry2018towards}, and the optimization-based C\&W attack~\cite{carlini2017towards}. 
For black-box attacks, we employed the most commonly used HSJA attack~\cite{chen2020hopskipjumpattack} and transfer-based attack, MI-FGSM~\cite{dong2018boosting} for comparative evaluation.
We consider an attack successful only when the input is misclassified into the specified target category~\cite{li2020towards}.

\noindent\textbf{Baselines.} We compare our method against three prevalent on-device model defense strategies: stochastic neural networks (SSNN~\cite{yu2021simple} and PNI~\cite{Xie_2019_CVPR}), adversarial training(A.T.)~\cite{madry2018towards}, and model architecture obfuscation (M.A.O.)~\cite{zhou2023modelobfuscator}.
We also compare the robustness and overall effectiveness of injecting noise directly into the raw weights~\cite{szentannai2020preventing} with \Name to clearly illustrate the superiority of our orthogonal decomposition strategy.

\noindent\textbf{Evaluation metrics.}
We quantify cross-node robustness using the attack success rate ($C\text{-}ASR$) and its complementary defense measure ($Anti\text{-}ASR$), where 
$Anti\text{-}ASR = 100\% - C\text{-}ASR$.
Due to space constraints, we present only the key experiments in the main paper, while extended evaluation details are included in the supplementary material.

\subsection{Main Results}

We quantitatively evaluate our proposed method for effectively defending against cross-node attacks, aiming to reduce the ASR of adversarial samples generated on compromised edge devices. Our main experiments thoroughly evaluate the effectiveness of \Name in defending against diverse cross-node attacks under various attack strengths, algorithms and knowledge of the adversaries.

\begin{figure*}[t]
    \centering
    \begin{minipage}[t]{0.49\textwidth}
        \centering
        \vspace{0pt}
        \includegraphics[height=4cm,width=\textwidth,keepaspectratio]{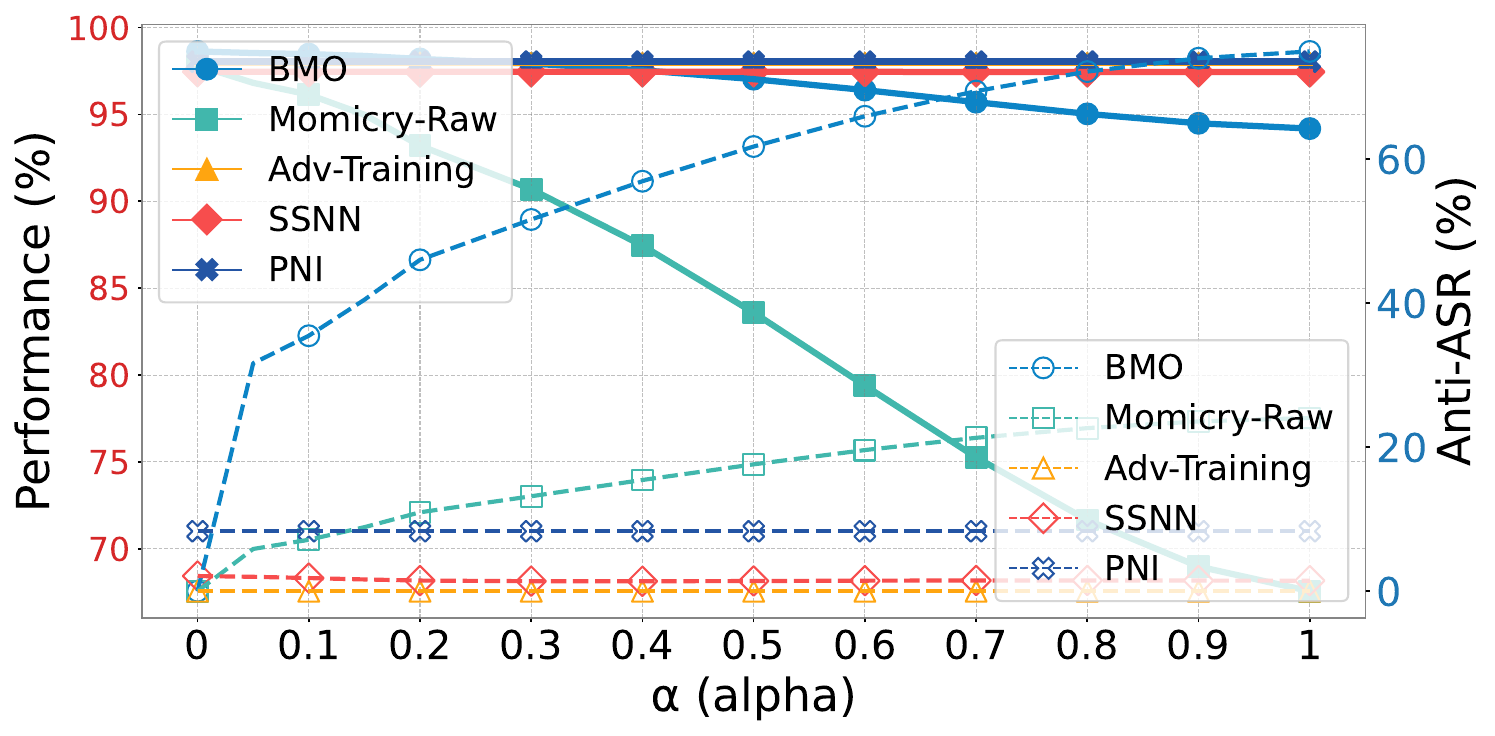}
        \caption{\textbf{Impact of model obfuscation on performance.}}
        \label{fig:performance2}
    \end{minipage}
    \begin{minipage}[t]{0.49\textwidth}
        \centering
        \vspace{0pt}
        \includegraphics[height=4cm,width=\textwidth,keepaspectratio]{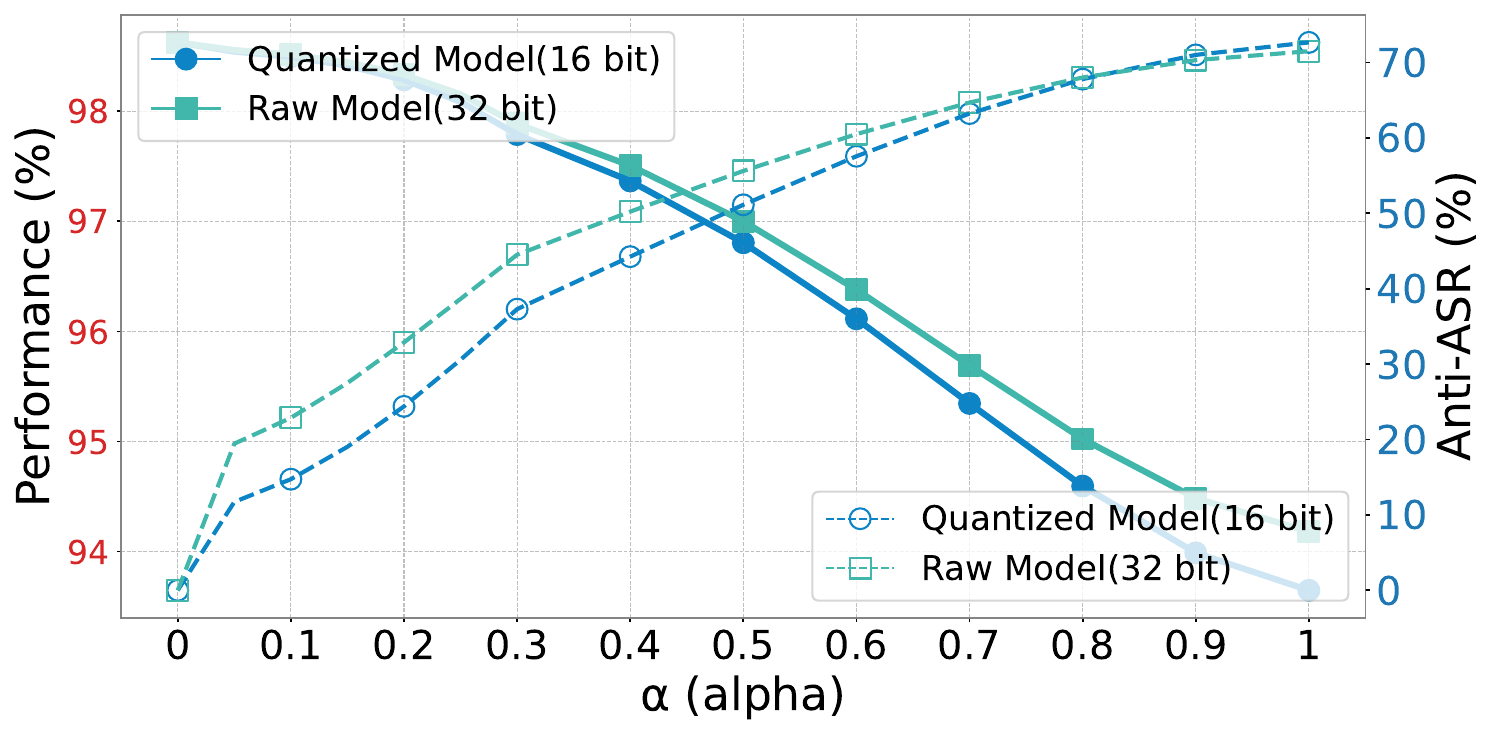}
        \caption{\textbf{Robustness against model quantization.}}
        \label{fig:Quantization}
    \end{minipage}
    \vspace{-10pt}
\end{figure*}

\noindent\textbf{Comparison with baselines.}
To validate the effectiveness of our defense in practical adversarial environments, we first compare \Name with representative baselines. This experiment aims to examine whether the proposed weight-obfuscation mechanism provides more consistent robustness than existing defenses when nodes that deploy the same models are compromised. 
Specifically, we include adversarial training, the raw weight modification approach~\cite{szentannai2020preventing}, SSNN~\cite{yang2023simple}, PNI~\cite{he2019parametric}, and model architecture obfuscation~\cite{zhou2023modelobfuscator} under various perturbation levels. 
The perturbation budget is constrained to $0.1$ to preserve perceptual consistency between the original and perturbed samples. This upper bound ensures that the injected perturbations remain visually imperceptible to human observers, maintaining the semantic integrity of the image content. A larger perturbation magnitude would introduce excessive pixel-level noise, which not only significantly degrades image fidelity but also alters high-level semantics, leading to unfair comparisons across models and methods~\cite{xiao2021improving}.

As shown in Figure~\ref{fig:main}\subref{fig:epsilon}, \Name consistently outperforms all defenses across the entire perturbation range. It maintains an Anti-ASR above 65\% for $\epsilon \le 0.05$ where a threshold for stealthy attacks, and remains around 55\% at $\epsilon = 0.1$. In contrast, adversarial training collapses completely with 0\% Anti-ASR, confirming its inability to handle cross-node perturbations. Among weight-obfuscation defenses, \Name achieves the highest resilience, surpassing raw weight modification by an average of 54.74\% (up to 63.33\%), while SSNN and PNI quickly degrade once $\epsilon$ exceeds 0.05.

In summary, \Name maintains over 45\% superiority on average, demonstrating that its parameter-space regularization provides a more durable barrier against adversarial perturbations. These results collectively confirm that \Name’s defense strength clearly stems not from memorizing specific perturbations, but from structurally enhancing decision-boundary smoothness across nodes.

\noindent\textbf{Effect of adversary’s knowledge on cross-node attack.} 
To comprehensively evaluate the impact of adversarial knowledge on C-ASR of cross-node attacks, we conduct experiments under three knowledge conditions: 
\ding{182} Adversaries access the model on a specific node to attack other nodes; 
\ding{183} Adversaries possess full knowledge of the raw model to attack obfuscated models on the edge nodes; 
\ding{184} Adversaries have full knowledge of a model on a specific node but aim to attack the raw model. 
These configurations simulate real-world scenarios where adversaries may have varying levels of knowledge about the deployed defenses and different access to raw models or specific nodes.

Figure~\ref{fig:main}\subref{fig:strict_knowledge}\&\ref{fig:performance2} illustrate the significant impact of attackers' knowledge on defending against adversarial attack in cross-node scenarios. 
When attackers exploit the original model with stolen model parameters, Anti-ASR remains alarmingly much lower, even at maximum defense intensity ($\alpha = 1.0$), growing only to 45.99\%. However, when the attacker only has access to a specific node, for raw model, Anti-ASR achieves 64.63\% at $\alpha = 0.6$, and further surges to 91.64\%at $\alpha = 1.0$; for other edge models, the data is 75.96\% and 89.33\% respectively.
Our experiment results highlight the importance of protecting the raw model as expected, as its leakage can significantly increase the overall threat of cross-node attacks across all nodes.

\noindent\textbf{Effectiveness against popular adversarial attacks.}
As shown in Figure~\ref{fig:Methodology}, without protection, adversarial attacks on one node can be replicated on others, leading to widespread compromise. 
In order to evaluate our defenses against various attacks, we use three representative white-box adversarial attack algorithms: 
FGSM~\cite{goodfellow2014explaining}, PGD~\cite{madry2018towards}, C\&W~\cite{carlini2017towards}, the decision-based black-box method HSJA~\cite{chen2020hopskipjumpattack} and the transfer-based method MI-FGSM~\cite{dong2018boosting}. These attacks were chosen to cover a broad spectrum of adversarial strategies under varying obfuscation budgets ($\alpha$). 

As illustrated in Figure~\ref{fig:main}\subref{fig:strict_attack_method}, across all settings, \Name consistently improves the Anti-ASR of the cross-node attack as $\alpha$ increases, indicating higher defense strength. 
For FGSM and PGD, Anti-attack success rates increase to 81.29\% and 72.01\%, respectively, showing strong defense against gradient-based attacks. C\&W and HSJA are also substantially suppressed, with final Anti-ASR of 76.60\% and 88.62\%. For the optimal obfuscation range ($\alpha \in [0.4, 0.6]$) where perturbations are moderate, the overall Anti-ASR of \Name on four attacks is 60.98\%. 
Additionally, the results show that our method achieves an Anti-ASR of 69.67\% on the transfer-based attack MI-FGSM, which clearly indicates that that \Name is effective on other adversarial attacks beyond cross-node attacks, such as transfer-based black-box attacks.
These results confirm that \Name provides a consistent and robust defense, effectively mitigating both white-box and black-box attacks.

\noindent\textbf{Robustness against model quantization.}
Quantization is commonly used on edge devices to reduce model size and inference latency by representing weights and activations with lower‐precision formats. 
However, quantized models can exhibit substantially different adversarial vulnerabilities from their full‐precision counterparts. To assess whether our protection remains effective after 16-bit quantization, we systematically evaluate the quantized models against cross-node attacks with four algorithms: FGSM~\cite{goodfellow2014explaining}, PGD~\cite{madry2018towards}, C\&W~\cite{carlini2017towards}, and HSJA~\cite{chen2020hopskipjumpattack}.

Both models maintain comparable accuracy on benign samples, with the quantized model experiencing a decrease of less than 0.01\%. This suggests that 16-bit quantization does not significantly compromise the accuracy of obfuscated models.
More importantly, as shown in Figure~\ref{fig:Quantization}, when $\alpha \in [0.4, 0.6]$, the Anti-ASR of the quantized model is only 3-6\% lower than that of the full-precision model. For instance, at $\alpha=0.5$, the quantized model achieves 55.30\% Anti-ASR, compared to 58.34\% for the original model. Notably, when $\alpha$ increases further (\eg, $\alpha \geq 0.8$), the quantized model consistently maintains clean accuracy over 94.82\% and achieves higher Anti-ASR than the original model.
The high robustness of \Name to model compression and deployment enables it to be readily suited for use in real-world edge computing scenarios.

\begin{figure*}[t]
    \centering
    \begin{subfigure}{0.22\textwidth}
        \centering
        \includegraphics[height=3.8cm,keepaspectratio]{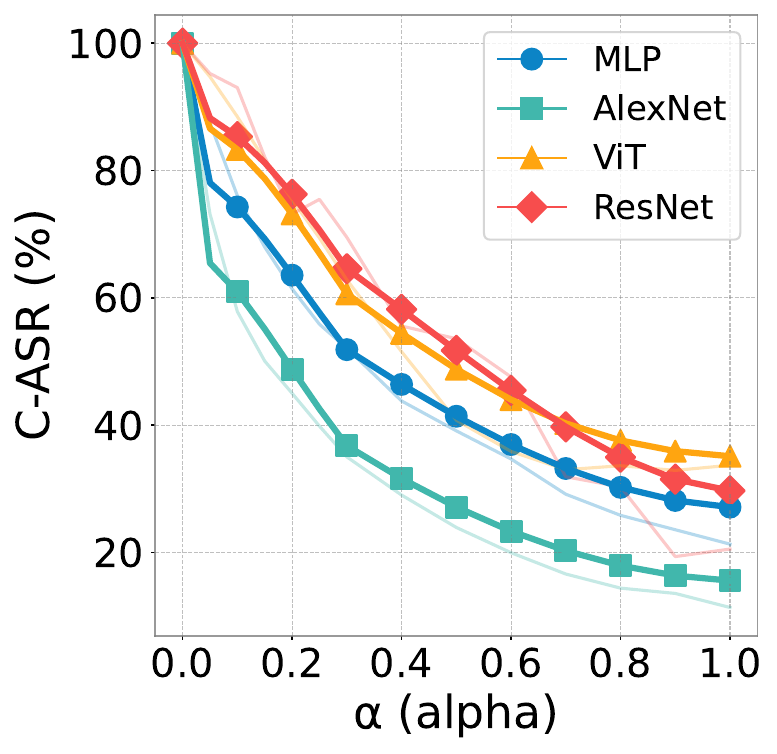}
        \caption{\textbf{Model Architecture}}
        \label{fig:strict_model_type}
    \end{subfigure}
    \hfill
    \begin{subfigure}{0.22\textwidth}
        \centering
        \includegraphics[height=3.8cm,keepaspectratio]{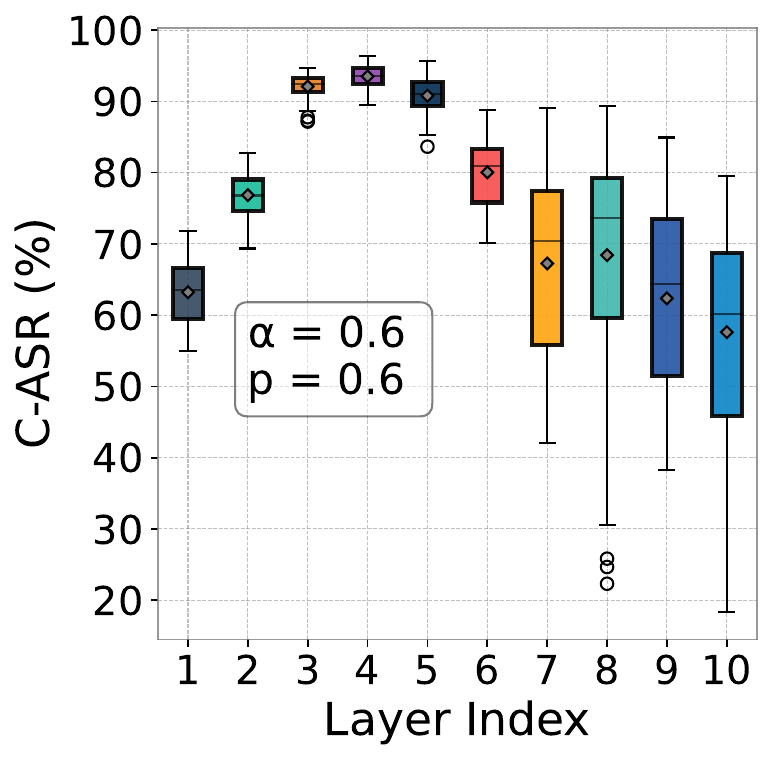}
        \caption{\textbf{Layer Indexing}}
        \label{fig:strict_layer}
    \end{subfigure}
    \hfill
    \begin{subfigure}{0.22\textwidth}
        \centering
        \includegraphics[height=3.8cm,keepaspectratio]{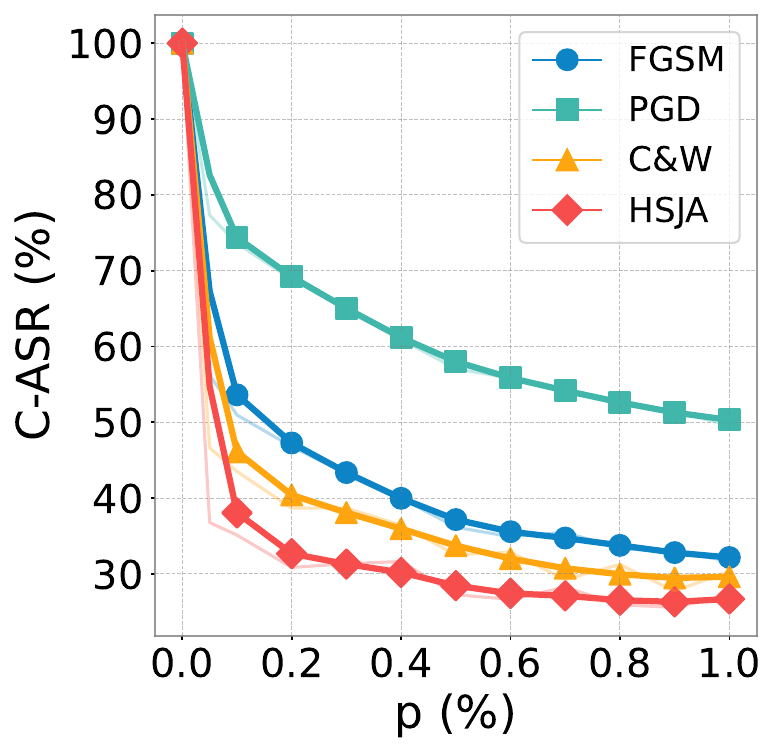}
        \caption{\textbf{Obfuscation Position}}
        \label{fig:strict_p}
    \end{subfigure}
    \hfill
    \begin{subfigure}{0.32\textwidth}
        \centering
        \includegraphics[height=3.8cm,keepaspectratio]{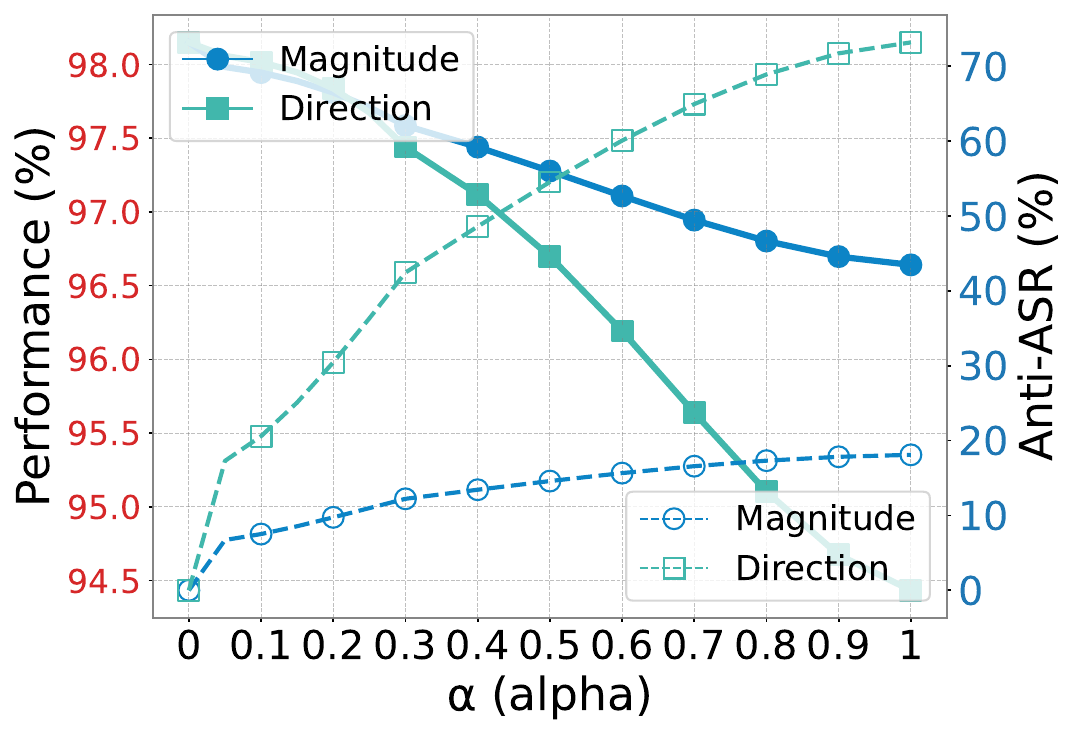}
        \caption{\textbf{Weight Component}}
        \label{fig:WeightComponent}
    \end{subfigure}
    \caption{\textbf{Ablation Studies of \Name.} Lower C-ASR and higher Anti-ASR indicate better defense performance against cross-node attacks, while higher Performance denotes better model performance on the target dataset.}
    \label{fig:Ablation}
    \vspace{-10pt}
\end{figure*}

\subsection{Ablation Studies}

In this section, we conduct ablation studies to verify the effectiveness of each hyperparameter of \Name, including obfuscation position, layer types and layer indexing. 
Additionally, we evaluate the defensive capability of perturbing magnitude components compared with direction components in our decomposition framework.

\noindent\textbf{Impact of obfuscation position selection.} 
The selection of which positions to perturb is critical for balancing robustness and efficiency in edge deployment. By strategically targeting only the most important directional components (controlled by parameter $p$), our method achieves strong defense with minimal modifications.
As shown in Figure~\ref{fig:Ablation}\subref{fig:strict_p}, the steepest drop occurs at very small $p$: for FGSM attacks, perturbing merely 10\% of weight directions ($p=0.1$) reduces the attack success rate from 100\% to 50.96\%, achieving more than half of the maximum possible protection. The marginal improvement becomes negligible beyond $p=0.5$, where the C-ASR drops to 36.17\% and only decreases by an additional 4.49\% when perturbing all weights ($p=1.0$). This pattern holds consistently across all attack methods, with average C-ASR reductions of 49.20\% from $p=0$ to $p=0.1$, 12.54\% from $p=0.1$ to $p=0.5$, and only 3.54\% from $p=0.5$ to $p=1.0$. This marginal benefit implies that perturbing a small proportion of high‑impact directions is sufficient to achieve most of the robustness benefit while maintaining strong resistance across attack types.

\noindent\textbf{Effect of model architecture.} 
Real-world deep learning systems typically employ diverse neural architectures, each exhibiting distinct vulnerability to adversarial attacks. Our experiments on three cornerstone architectures (MLPs, CNNs, and ViTs) validate the method's generalizability.
Figure~\ref{fig:Ablation}\subref{fig:strict_model_type} reveals that, for CNNs, the defense reduces C-ASR to 11.35\%, demonstrating its exceptional effectiveness that likely stems from the convolutional filters' localized sensitivity. 
MLPs exhibit more gentle but still significant improvements, with C-ASR declining steadily to 21.30\%, proving \Name's robustness and effectiveness in fully-connected networks where vulnerabilities are more evenly distributed. 
ViTs present the most challenging case, maintaining a 33.70\% C-ASR even at full defense strength, but over 65\% relative improvement still proves \Name's applicability even to sophisticated architectures. 
The results further demonstrate that \Name offers a foundational and broadly applicable defense for DL models.

\noindent\textbf{Role of layer indexing.} 
To investigate the relationship between layer depth and obfuscation effectiveness, we obfuscate weights at different layers on a 10-layer MLP model and measure their effect on resistance to cross-node attacks. 
As shown in Figure~\ref{fig:Ablation}\subref{fig:strict_layer}, shallow layers exhibit strong but relatively stable response to perturbations, with C-ASR ranging from 76.84\% to 63.21\%; middle layers function primarily as ``pass-through'' regions for adversarial gradients, showing minimal impact on final defense performance, showing relatively higher average C-ASR values from 93.51\% to 80.04\%; deep layers play a crucial role, also show hyper-sensitive response with a small $\alpha$ change, with an average C-ASR of 63.91\%, as they directly influence final predictions. This depth stratification suggests that adversarial attacks primarily originate in shallow layers but also manifest in deep layers, thereby indicating the critical importance of careful strategic layer selection.

\noindent\textbf{Weight component.}
To further evaluate the impact of weight decomposition in our defense, we inject noise into the magnitude components instead of the direction components.
As illustrated in Figure~\ref{fig:Ablation}\subref{fig:WeightComponent}, across 10 perturbation strengths, injecting noise into the magnitude component makes accuracy on clean data drop slightly, from 98.15\% to 96.13\%, but yields significantly weaker robustness under cross-node attacks by an average of 12.54\% and up to 19.51\% in the best case. 
In contrast, perturbing the direction component produces dramatic Anti-ASR improvement by an average of 44.98\% and up to 81.89\%, at a maximum loss in clean accuracy with a 6.55\% drop.
This indicates that magnitude perturbations cannot effectively obfuscate the model’s decision boundary,  while the direction matrix governs the geometry of the model’s decision boundaries, making it critical for our overall defense strategy.

\section{Conclusion}

We initiate the study of safety risks in edge systems under cross-node attacks, where adversaries exploit a compromised node to launch white-box attacks on others. To counter this threat, we propose \Name, a weight-obfuscation defense that produces device-specific models with distinct decision boundaries, thereby reducing adversarial transferability across nodes.
We conduct extensive evaluations of \Name on three popular on-device model architectures using five popular adversarial attack methods. The results show that \Name significantly outperforms the strongest baseline in defending against cross-node attacks.
Our research further highlights the effectiveness of \Name as a practical and scalable defense against cross-node attacks. 

\clearpage
{
    \small
    \bibliographystyle{ieeenat_fullname}
    \bibliography{main}
}

\clearpage
\maketitlesupplementary
\appendix

\begin{figure}
    \centering
    \includegraphics[width=\linewidth]{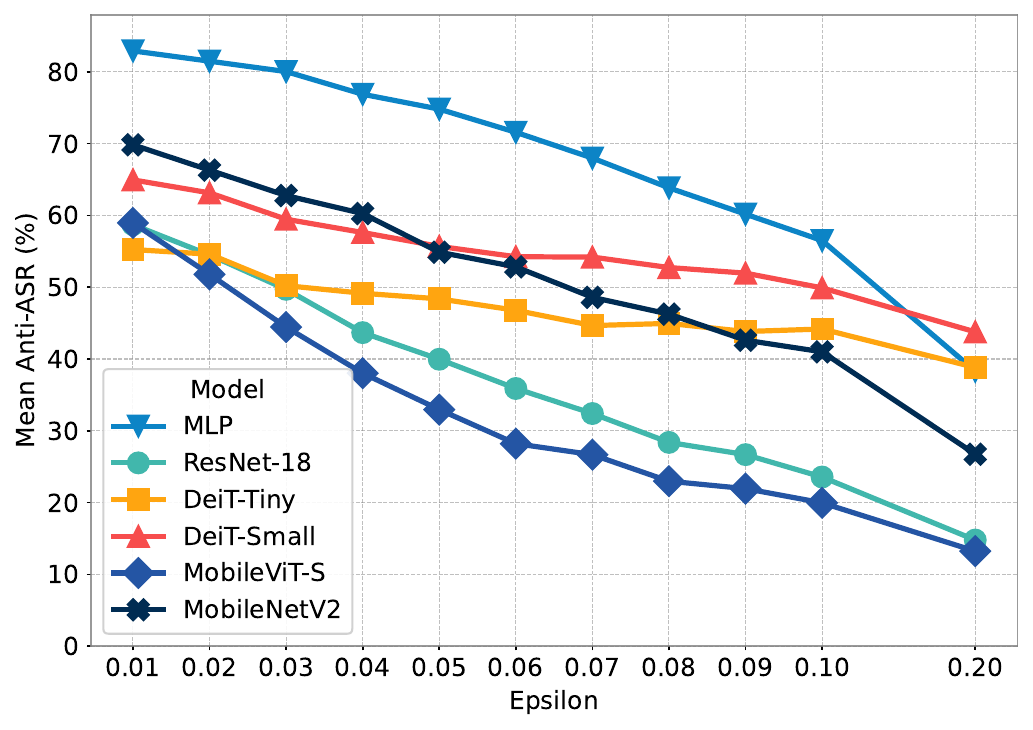}
    \caption{\textbf{Mean Anti-ASR under increasing perturbation budgets across six representative edge-device models.}}
    \label{fig:models}
\end{figure}

\begin{table*}[h]

  \centering
  \caption{\textbf{Cross-Dataset and Cross-Architecture Robustness (Anti-ASR) Across Six Edge-Device Models}}
    \begin{tabular}{lcccccc}
    \toprule
    Anti-ASR & \multicolumn{6}{c}{Model}    \\
    \cmidrule(lr){2-7}
    ($\epsilon = 0.05$) & MLP   & ResNet & DeiT\_tiny & DeiT\_small & MobileViT\_s & MobileNet\_v2 \\
    \midrule
    MNIST & 86.70/2.43 & 69.21/2.96 & 72.20/2.08 & 76.39/3.04 & 61.71/2.80 & 89.76/2.99 \\
    CIFAR10 & -     & 40.06/2.78 & 35.16/2.06 & 47.04/2.85 & 61.22/2.75 & 61.01/2.86 \\
    CIFAR100 & -     & 52.72/2.66 & 46.34/3.15 & 56.23/3.19 & 45.52/2.78 & 58.73/2.93 \\
    ImageNet & -     & 40.55/3.13 & 50.53/3.25 & 62.04/2.99 & 38.72/2.85 & 53.49/2.99 \\
    \bottomrule
    \end{tabular}%
  \label{tab:dataset}%
\end{table*}%

\begin{figure*}[t]
    \centering

    \begin{subfigure}[t]{0.45\textwidth}
        \centering
        \includegraphics[width=\linewidth]{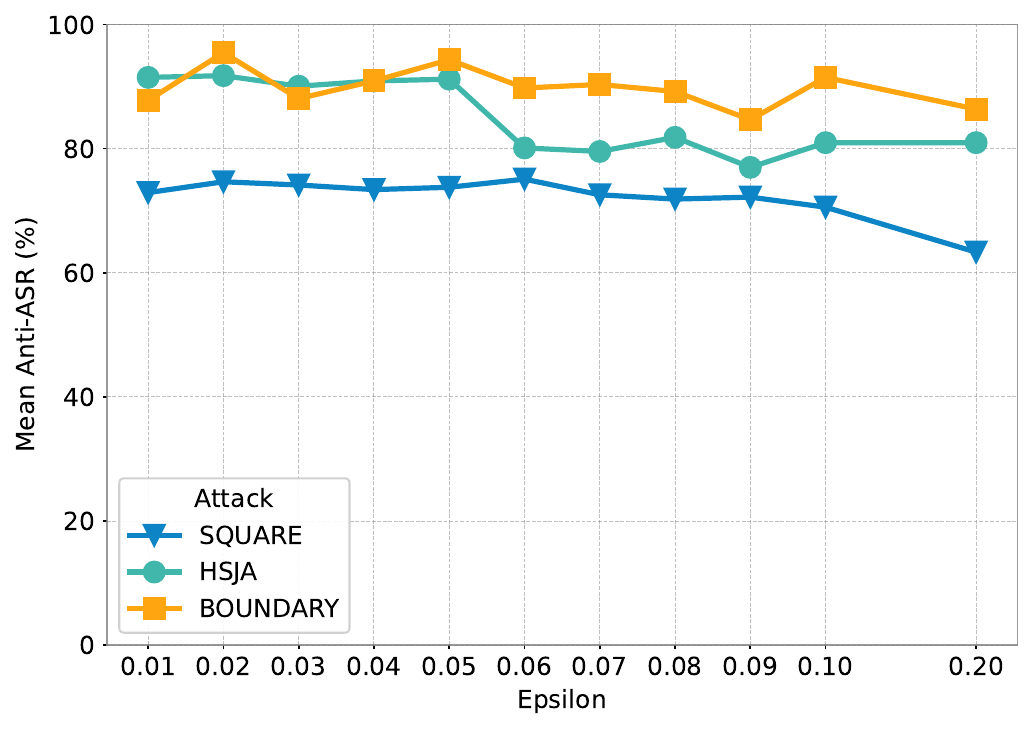}
        \caption{\textbf{Query-based}}
        \label{fig:Query_based}
    \end{subfigure}
    \hfill
    \begin{subfigure}[t]{0.45\textwidth}
        \centering
        \includegraphics[width=\linewidth]{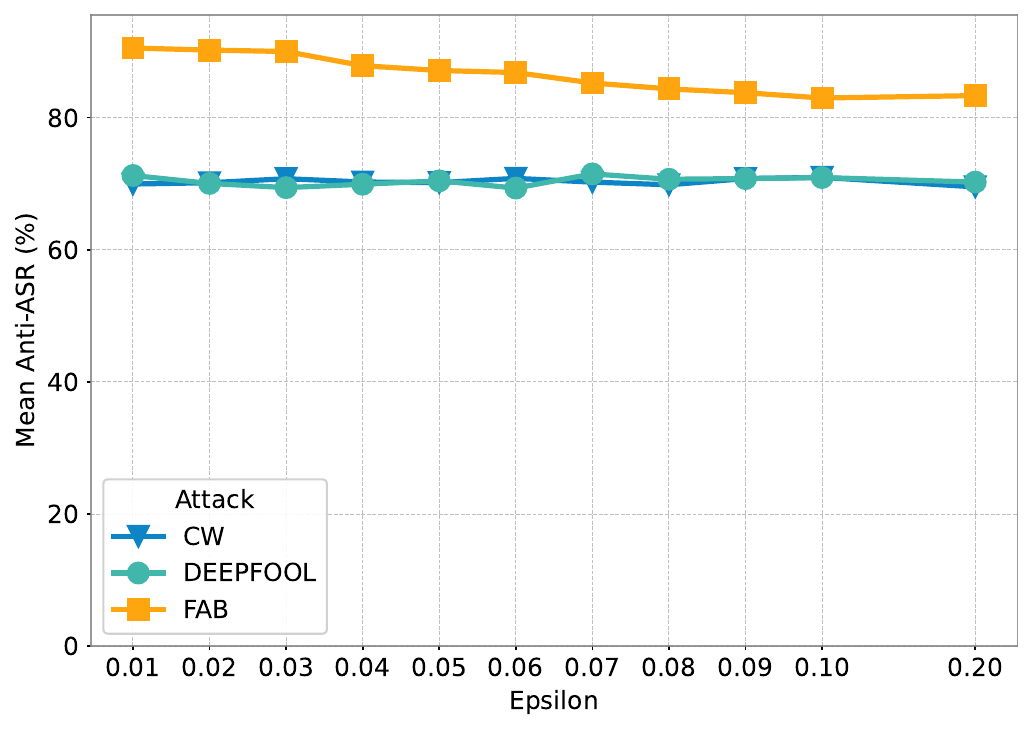}
        \caption{\textbf{Optimization-based}}
        \label{fig:Optimization_based}
    \end{subfigure}

    \vspace{0.3cm}

    \begin{subfigure}[t]{0.45\textwidth}
        \centering
        \includegraphics[width=\linewidth]{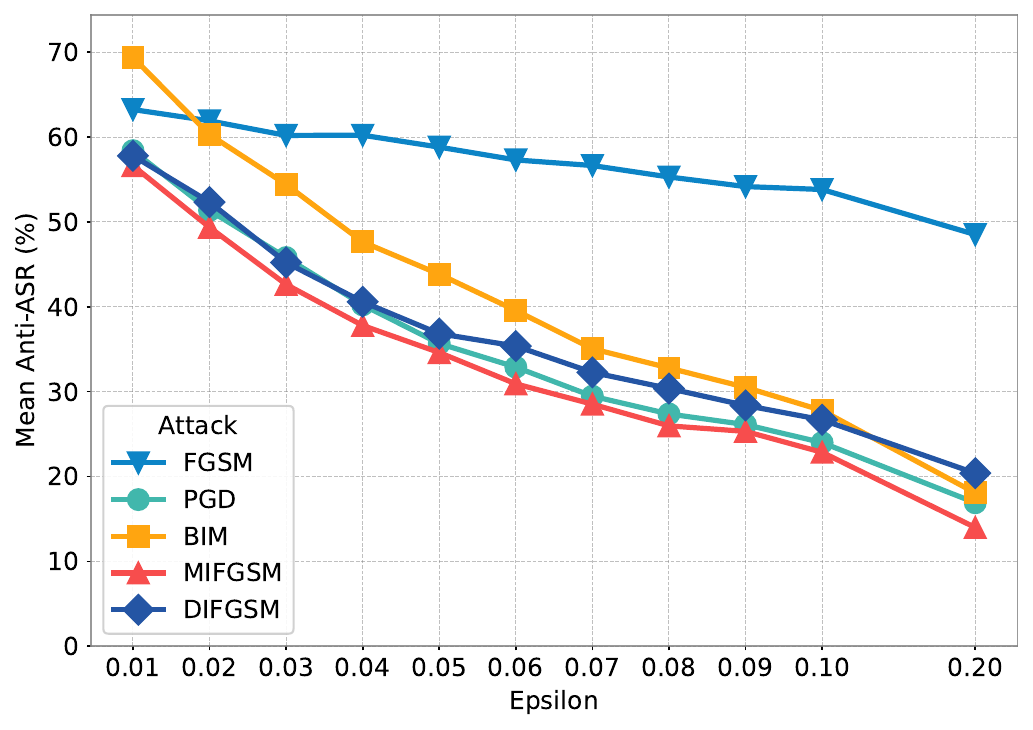}
        \caption{\textbf{Gradient-based}}
        \label{fig:Gradient_based}
    \end{subfigure}
    \hfill
    \begin{subfigure}[t]{0.45\textwidth}
        \centering
        \includegraphics[width=\linewidth]{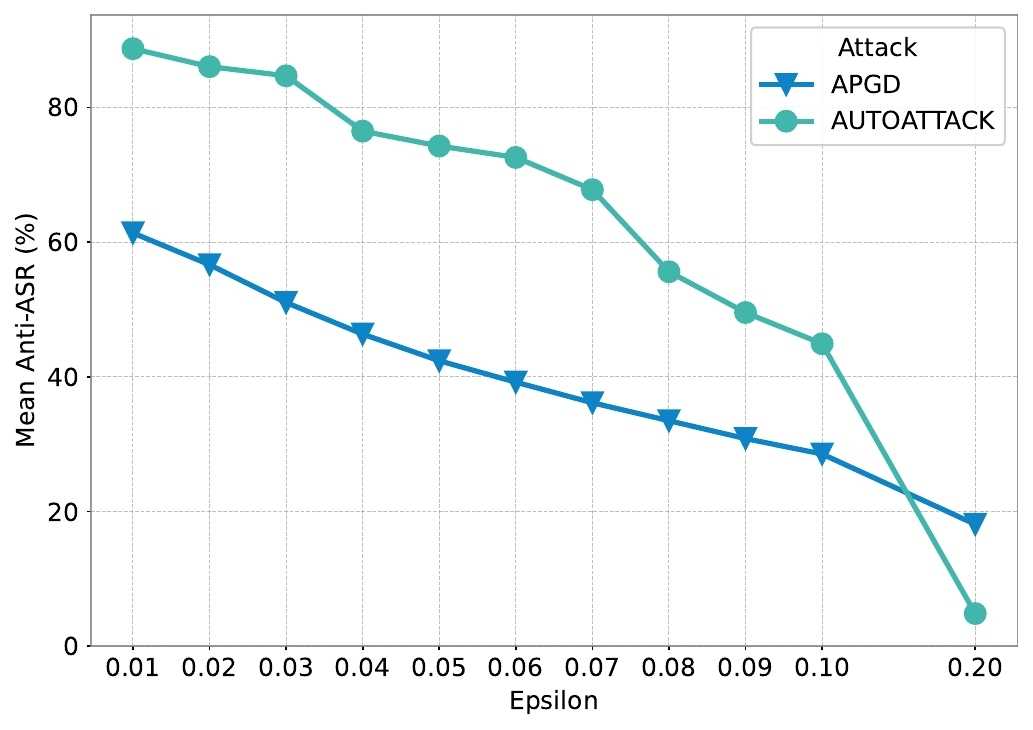}
        \caption{\textbf{AutoML-based}}
        \label{fig:AutoML_based}
    \end{subfigure}

    \caption{\textbf{Effectiveness Against Different Attack Families.}}
    \label{fig:attack_methods}
\end{figure*}

\section{Evaluation on Diverse DNN Architectures}
Existing studies~\cite{singh2023edge,oliynyk2023know} have shown that four major model families form the dominant architectures deployed on edge devices, namely MLPs, CNNs, Vision Transformers, and lightweight mobile architectures. Thus, we employ these six representative models from each family to comprehensively evaluate the stability of BMO across diverse edge-device architectures, including MLP, ResNet, DeiT\_small, DeiT\_tiny, MobileNet and MobilViT. 

Additionally, we conduct extensive experiments under a wide range of perturbation budgets.
Prior work has shown that strong adversarial perturbations inevitably reduce the stealthiness of the attack, and perturbations greater than $\epsilon > 0.05$ can even be visually distinguishable to human observers~\cite{veerabadran2023subtle}. Such visible distortions fall outside standard threat models for evaluating practical adversarial robustness. Therefore, our experiments investigate the performance in the realistic and imperceptible region of $\epsilon \le 0.05$. 

As shown in Figure~\ref{fig:models}, our method consistently maintains a strong low-budget robustness: when the attack budget is $\epsilon =  0.05$, all the model achieves more than 30\% average Anti-ASR on 4 datasets and 13 adversarial attacks, and the best-performing architectures exceed 78\%. As $\epsilon$ increases, all models exhibit the expected degradation, but the decay rate varies by architectural family. Transformer-based models (\ie, DeiT-Tiny/Small) exhibit the most stable robustness profiles, maintaining roughly 45–55\% Anti-ASR even at $\epsilon = 0.10$, while CNN-based architectures and MobileViT-S show moderately lower, but still meaningful, robustness under the same budget. Lightweight models such as MLP and MobileNetV2 also retain non-trivial Anti-ASR, though their effectiveness begins to decrease earlier as $\epsilon$ increases. Importantly, the overall decline across all models is gradual rather than abrupt, indicating that the defense remains consistently effective across heterogeneous architectures. These results suggest that while the absolute robustness levels vary with architectural design, our method provides \textbf{reliable and architecture-agnostic protection}, delivering consistent defensive benefits across the diverse model families commonly deployed on edge devices.

\section{Performance Across Different Datasets}

To evaluate the capability of our method in tackling diverse datasets, we conduct experiments across four popular datasets widely applied in image classification benchmarks, including MNIST~\cite{deng2012mnist}, CIFAR-10~\cite{krizhevsky2009learning}, CIFAR-100~\cite{krizhevsky2009learning}, and ImageNet~\cite{deng2009imagenet}.
These datasets cover different input dimensionalities, label granularities, and distributional complexities, providing a comprehensive evaluation of our BMO in generalizing diverse datasets. To explore the performance of BMO under diverse datasets, we employ the most powerful adversarial attack APGD~\cite{croce2020reliable} in evaluation. This ensures a consistent and rigorous robustness comparison across these datasets, while evaluation in surviving multiple adversarial attacks is presented in Appendix~\ref{app:attack_families}.

As shown in Table~\ref{tab:dataset}, we report each model's robustness against APGD under an attack budget of $\epsilon = 0.02$ and a defense constraint of $\Delta\mathrm{ACC}<3.5\%$. 
This setting follows prior works indicating that perturbations at $\epsilon = 0.02$ remain imperceptible while still inducing strong adversarial effects, and constraining $\Delta\mathrm{ACC}<3.5\%$ ensures that the defense maintains acceptable predictive performance for on-device deployment.
Across all settings, BMO maintains consistently strong Anti-ASR, exhibiting no reliance on model-specific inductive biases and remaining stable across tasks ranging from low-dimensional digit recognition (MNIST) to high-resolution natural image classification (ImageNet). 
Despite substantial distributional differences across datasets, including variations in label granularity, visual diversity, and input dimensionality, its robustness degradation remains smooth, bounded, and predictable. 
These results indicate that BMO \textbf{generalizes across both architectural paradigms and dataset regimes well}, providing a reliable and transferable resilience that extends beyond specific model or task configurations. 

\section{Robustness against Popular Adversarial Attacks}\label{app:attack_families}

To assess the performance of BMO under diverse adversarial techniques, we evaluate its robustness against four major types of adversarial attacks, following the taxonomy widely adopted in recent studies~\cite{akhtar2018threat,wang2022triangle,hamm2017machine}.
In evaluation, our results shown in Figure~\ref{fig:attack_methods}\subref{fig:Query_based}–\subref{fig:AutoML_based} reveal that BMO exhibits stable robustness against optimization-based and query-based attacks across a wide range of perturbation budgets, while also providing reliable protection against gradient-based and autoML-based attacks with imperceptible perturbation ($\epsilon<0.05$).

As shown in Figure~\ref{fig:attack_methods}\subref{fig:Query_based}\&\subref{fig:Optimization_based}, BMO exhibits the powerful robustness against optimization-based and query-based adversarial attacks, showing only minimal sensitivity to increasing perturbation budgets. This stability arises because these attacks rely heavily on local decision-boundary geometry or iterative query refinement. The representation-level randomness introduced by BMO disrupts these fine-grained geometric cues, making it difficult for such attacks to reuse boundary features from exploited nodes. As a result, their effectiveness remains limited even when the perturbation magnitude increases. Empirically, BMO maintains approximately 65–70\% Anti-ASR under C\&W and DeepFool across all budgets and achieves over 85\% Anti-ASR under FAB, confirming its resistance to attacks that depend on precise boundary manipulations.
Query-based attacks follow a similar trend: BMO sustains 70–75\% Anti-ASR under Square Attack and preserves exceptionally high robustness under HSJA and BoundaryAttack ($\approx$90\% at $\epsilon \le 0.03$ and $\approx$80\% at $\epsilon \le 0.10$), demonstrating the notably limited effectiveness of black-box boundary-search methods under our method.

In contrast, gradient-based and autoML-based attacks exert stronger challenges on BMO as the perturbation budget grows. 
Unlike boundary dependent methods, these attacks optimize class-level gradients or adaptively adjust their optimization trajectory, allowing them to bypass the local boundary geometry that BMO obfuscated. Consequently, once the perturbation becomes sufficiently large, the adversarial perturbations disrupt the decision boundary entirely, weakening BMO's defensive effect. Nevertheless, BMO remains highly effective within the imperceptible adversarial perturbation regime ($\epsilon \le 0.05$), consistently mitigating stealthy adversarial attacks that fall within realistic threat models. In this range, as shown in Figure~\ref{fig:attack_methods}\subref{fig:Gradient_based}\&\subref{fig:AutoML_based}, BMO substantially suppresses the effectiveness of gradient-based and autoML-based attacks, as the limited perturbation budget confines their adversarial examples to regions close to the decision boundary. 
Under such restricted budgets, the obfuscation-induced perturbation of the boundary makes the gradient information extracted from compromised nodes highly inconsistent across models. 
Consequently, adversarial inputs crafted from these gradients fail to transfer to other nodes, further limiting the effectiveness of gradient-based and autoML-based attacks in this regime.
These results indicate that BMO provides reliable protection against stealthy adversarial examples, which constitute the primary security concern in real-world deployments.

When the perturbation budget increases beyond this limitation ($\epsilon > 0.05$), the injected noise begins to visibly distort the image, and adversarial examples cross and move far beyond the decision boundary, falling outside standard threat models. 
In these over-perturbed settings, the injected adversarial noise introduces clear visual artifacts and severely distorts the input. 
Such severe distortions drive the adversarial examples far beyond the decision boundary, thereby diminishing the influence of BMO.
This reduction arises from the strong adversarial artifacts introduced by the large perturbation, where the attacks themselves lose stealth and fail to keep aligned with the intended threat models.

Together, these observations show that BMO delivers \textbf{strong and consistently stable robustness against optimization- and query-based attacks across a wide range of perturbation budgets}, while retaining \textbf{reliable defensive capability against gradient- and autoML-based attacks within imperceptible perturbation levels}. This demonstrates the broad applicability of our method across fundamentally different adversarial regimes.

\end{document}